\documentclass{article}
\usepackage{spconf,amsmath,graphicx, url}
\usepackage{amssymb, bm}
\usepackage{pifont}
\newcommand{\cmark}{\ding{51}}%
\newcommand{\xmark}{\ding{55}}%
\ninept

\title{Endpoint Detection for Streaming End-to-End Multi-talker ASR}
\name{Liang Lu$^\dagger$\thanks{The work was done at Microsoft.}, Jinyu Li$^\ddagger$ and Yifan Gong$^\ddagger$}
\address{$^\dagger$ Otter.ai, Mountain View, CA, USA \\
$^\ddagger$Microsoft Corp. USA\\}

\begin{document}
\ninept
\maketitle
\begin{abstract}
\end{abstract}
Streaming end-to-end multi-talker speech recognition aims at transcribing the overlapped speech from conversations or meetings with an all-neural model in a streaming fashion, which is fundamentally different from a modular-based approach that usually cascades the speech separation and the speech recognition models trained independently. Previously, we proposed the Streaming Unmixing and Recognition Transducer (SURT) model based on recurrent neural network transducer (RNN-T) for this problem and presented promising results. However, for real applications, the speech recognition system is also required to determine the timestamp when a speaker finishes speaking for prompt system response. This problem, known as endpoint (EP) detection, has not been studied previously for multi-talker end-to-end models. In this work, we address the EP detection problem in the SURT framework by introducing an end-of-sentence token  as an output unit, following the practice of single-talker end-to-end models. Furthermore, we also present a latency penalty approach that can significantly cut down the EP detection latency. Our experimental results based on the 2-speaker LibrispeechMix dataset show that the SURT model can achieve promising EP detection without significantly degradation of the recognition accuracy.  

\section{Introduction}
\label{sec:intro}

Multi-talker automatic speech recognition (ASR) refers to the problem of transcribing (partially) overlapped speech from multiple speakers in a conversion or a meeting scenario. Conventional approaches to deal with this challenge are usually based on the modularized framework, e.g., cascading the speech separation and speech recognition modules that are trained independently~\cite{chen2020continuous, yoshioka2018recognizing}. While working reasonably well in practice, these approaches have the drawback of high overall system complexity. More importantly, optimizing each module of the cascaded system does not guarantee to improve the overall system performance. Motivated by its success in single-talker speech recognition, recently, researchers have been exploring end-to-end (E2E) based approaches to address the challenges in the multi-talker scenario with a single neural network model that all the parameters can be trained jointly.

In the literature, most existing E2E-based multi-talker speech recognition systems are based on the sequence-to-sequence (S2S) with attention models~\cite{chang2019end, settle2018end, kanda2020serialized}. While promising results are achieved, these models are constrained to be offline only, and they usually fail when being applied directly to long-form audios of meetings or conversations without prior audio segmentation due to the offline nature of these models. Consequently, for a scenario that requires streaming speech transcription, i.e., the system is required to show transcriptions when speakers are still speaking, these approaches are not very applicable.  Inspired by an offline RNN-T~\cite{graves2012sequence} based multi-talker model~\cite{tripathi2020end}, we recently proposed the Streaming Unmixing and Recognition Transducer (SURT) model~\cite{lu2020streaming} to address the limitations of existing solutions. In~\cite{lu2020streaming}, we showed that SURT can achieve promising recognition accuracy for 2-speaker overlapped scenario, similar to the observation of a concurrent work~\cite{sklyar2020streaming}.  In~\cite{raj2021continuous}, we further demonstrated that SURT can directly transcribe continuous long-form audio without prior audio segmentation, and achieved competitive recognition accuracy compared with offline E2E models.   

However, to be deployed to real application scenarios, streaming E2E models are usually required to determine the timestamp when the speaker finishes speaking, i.e., EP detection. Accurate EP detection is the key for prompt system response and the overall user experience. While there have been a few studies in EP detection for single-talker E2E ASR~\cite{chang2019joint}, to the best our knowledge, there is no prior work on EP detection for the E2E multi-talker ASR problem. In this work, we address this problem in the SURT framework. In literature, conventional EP detection is usually achieved by Voice Activity Detection (VAD)~\cite{hariharan2001robust, thomas2015improvements, eyben2013real} that classifies the acoustic features into silence and speech. However, as argued in~\cite{chang2019joint}, the VAD-based approach solely relies on the acoustic information while ignoring the cues form the language model. More importantly, the conventional VAD-based approach cannot handle overlapped speech in a multi-talker scenario. Motivated by~\cite{chang2019joint}, we study a simple approach that introduces an additional end-of-sentence ($\langle \text{eos} \rangle$) token to the output of SURT model for EP detection. During inference, we use the timestamp that the model emits the $\langle \text{eos} \rangle$ token as the EP timestamp. To reduce the EP detection latency, we also study a latency penalty approach for model training. From our experiments on LibrispeechMix dataset, this approach enjoys low engineering complexity and negligible computational overhead, while achieving promising EP detection performance. 

The contributions of this paper are summarized as follows. 
\begin{itemize}
 \item We perform the first study of the EP detection in streaming E2E multi-talker ASR.
 \item We present a latency penalty approach that can significantly cut down the latency of EP detection based on the $\langle \text{eos} \rangle$ token.
 \item We extend the previous RNN-T based streaming E2E multi-talker ASR with a Transformer-Transducer (T-T) ~\cite{zhang2020transformer, yeh2019transformer}. 
\end{itemize}


\section{Transducer-based ASR Model}
\label{sec:T}

Neural transducer models such as RNN-T and T-T directly compute the sequence-level conditional probability of the label sequence given the acoustic features by marginalizing over all the possible alignments. Unlike S2S models, neural transducers perform sequence transduction in a time-synchronous fashion, and hence are suitable for streaming speech recognition. A neural transducer model is usually composed of three components, an audio encoder $f(\cdot)$, a label prediction network $g(\cdot)$ and a joint network $j(\cdot)$. Given an acoustic feature sequence $X=\{x_1, \cdots, x_T\}$ and its corresponding label sequence $Y=\{y_1, \cdots, y_U\}$, where $T$ is the length of the acoustic sequence, and $U$ is the length of the label sequence, the feature representations can be obtained as
\begin{align*}
\bm{h}_{1:t}^f = f(x_{1:t}), \hskip4mm \bm{h}_{1:u}^g = g(y_{0:u-1}),
\end{align*}
where $1 \le t \le T, 1 \le u \le U$ and $y_0$ denotes the blank token. The two feature representations are fused together by a joint network $j(\cdot)$ as
\begin{align}
\label{eq:joint}
\bm{z}_{t, u} = j({\bm h}_t^f, \bm{h}_u^g),
\end{align}
and the conditional probability of the token $y_u$ is obtained by a Softmax function, i.e., 
\begin{align}
P(y_u \mid \bm{z}_{t, u}) = \text{Softmax}(\bm{W}\bm{z}_{t,u} + \bm{b}),
\end{align}
where $\bm{W}$ and $\bm{b}$ are the weight matrix and bias. Given $P(y_u \mid \bm{z}_{t, u})$, the sequence-level conditional probability can be computed by dynamic programming, and the RNN-T loss can be defined as the negative log-likelihood as:
\begin{align}
\label{eq:nll}
\mathcal{L_{\text{rnnt}}} (Y, X) = -\log P(Y \mid X).
\end{align}
Details about the loss function can be found in~\cite{graves2012sequence}. 

\subsection{Transformer Transducer}
\label{sec:TT}

Previously, streaming E2E models usually adopt the RNN-T model structure, which employs the long short-term memory (LSTM) networks for both the audio encoder $f(\cdot)$ and the prediction network $g(\cdot)$. Recently, Transformers have been proven to be competitive for speech recognition problems~\cite{karita2019comparative}. As a result, T-T was proposed to replace the LSTM network by a Transformer network for the audio encoder $f(\cdot)$ and similarly for the prediction network. While there are a few different implementations of T-T, in this work, we closely follow the T-T model architecture presented in~\cite{chen2021developing}. In particular, we still use LSTMs for the prediction network in the T-T model for the sake of inference speed, while the audio encoder is based a Transformer network.

\begin{figure}[t]
\small
\centerline{\includegraphics[width=0.4\textwidth]{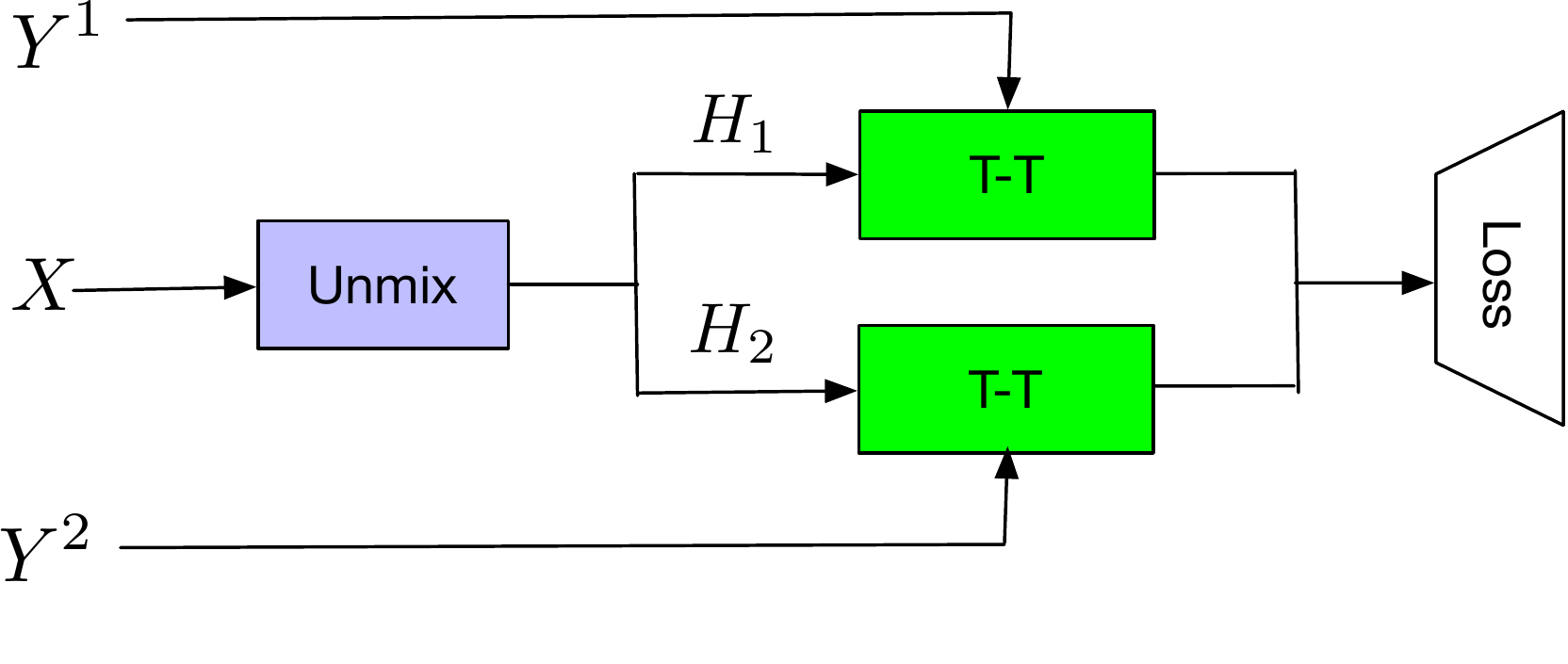}}
\vskip-4mm
\caption{T-T based SURT model. The unmixing module maps the input audio into two channels, and the output of each channel is processed by a T-T based recognition network. Note that the T-T model parameters are shared for both channels. }  
\label{fig:surt}
\vskip -2mm
\end{figure}

\section{SURT: Review}
\label{sec:surt}

SURT~\cite{lu2020streaming} is composed of two key components, the {\it unmixing} module and the {\it recognition} module. An example of an SURT model that is based on T-T is shown in Figure~\ref{fig:surt}. The role of the unmixing module is similar to speech separation. It takes the audio feature sequence $X$ that may have an {\it arbitrary} degree of overlapped speech from two speakers as input, and then transforms the feature representations into two channels. In each channel, we assume that the feature representation is only from one speaker at each time step. The reason to use only two output channels is that in most real multi-talker scenarios, there are at most two active speakers at each time step. Given the output from the unmixing module, the transducer-based recognition module is then applied to map the feature sequence into their corresponding label sequence. Different from the standard speech separation, the unmixing module is not trained by an independent signal reconstruction loss. Instead, both of the unmixing and recognition modules in SURT are trained jointly using an ASR loss as explained below. 

In~\cite{lu2020streaming}, we compared two network architectures for the unmixing module. In this work, we focus on the mask-based structure with convolutional neural networks (CNNs) due to its simplicity and performance. Specifically, the unmixing module has two CNN-based encoders that transform the audio feature input $X$ into two hidden feature representations $H_1$ and $H_2$ as:
\begin{align}
 \begin{array}{ll}
 M =  \sigma(\text{CNN}_\text{mask}(X)),  & \bar{X} = \text{CNN}_\text{enc}(X), \\
H_1 = \bar{X}* M,  & H_2 = \bar{X} - H_1,
 \end{array}
\end{align}
where $\sigma$ denotes the Sigmoid function, $M$ is a mask with each element ranging from 0 to 1, and $\text{CNN}_\text{mask}$ and $\text{CNN}_\text{enc}$ are two CNN encoders. The two feature representations are then fed into the same recognition module to compute the corresponding RNN-T losses using the two label sequences $Y^1$ and $Y^2$. Without any constraint, however, it is unclear the correspondence between the feature sequences $(H_1, H_2)$ and label sequences $(Y_1, Y_2)$, known as the label permutation problem. A widely used approach to address this problem is Permutation Invariant Training (PIT)~\cite{yu2017permutation}, which considers all the possible label assignments. While being flexible and effective, PIT has the drawback of high computational complexity, which is in the order of $O(N!)$ for the mixed signal with $N$ speakers. In~\cite{lu2020streaming}, we showed that a simpler label assignment strategy, referred as Heuristic Error Assignment Training (HEAT), can achieve comparable recognition accuracy for SURT models with the computational complexity in the order of $O(N)$. Unlike PIT, HEAT only considers one possible label assignment based on some heuristic information, for example, the spoken order of the utterances according to their starting times. Suppose $Y^1$ is the first spoken utterance, HEAT defines the loss function simply as
\begin{align}
\label{eq:lp}
\mathcal{L}_{\text{asr}}(X, Y^1, Y^2) = \mathcal{L}_{\text{rnnt}}(Y^1, H_1) + \mathcal{L}_{\text{rnnt}}(Y^2, H_2),
\end{align}
which means that the first channel and its corresponding feature representation $H_1$ are always assigned to the first-come speaker of an audio mixture. Likewise, the second channel is always dedicated to the other speaker. The HEAT loss function can drive the network to learn the corresponding channel assignments. More detailed comparison and analysis of HEAT vs. PIT are given  in~\cite{lu2020streaming, raj2021continuous}.

\begin{figure}[t]
\small
\centerline{\includegraphics[width=0.45\textwidth]{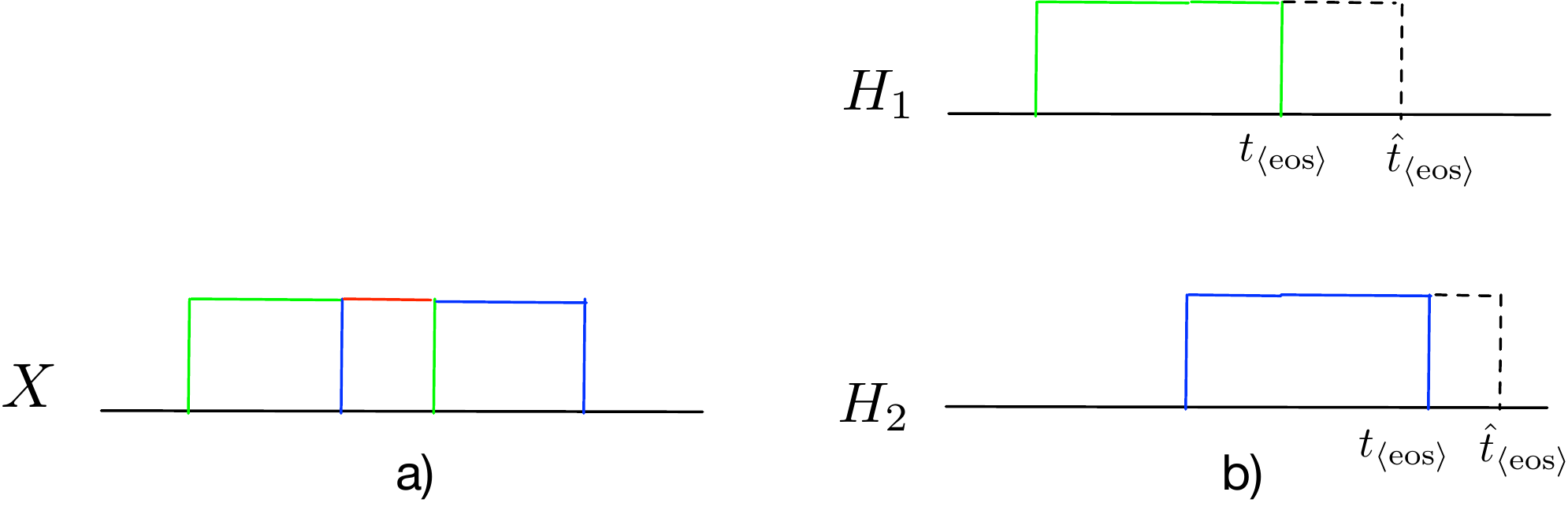}}
\vskip-2mm
\caption{a) An example of a mixture audio signal $X$ with speech from two speakers, denoted as green and blue respectively. b) After the unmixing module, the feature representations are mapped into two channels, denoted as $H_1$ and $H_2$ respectively. Note that $X$ is in the feature space, while $H_1$ and $H_2$ are hidden vector representations. $t_{\langle \text{eos} \rangle}$ is the ground-truth EP timestamp, while $\hat{t}_{\langle \text{eos} \rangle}$ denotes the predicted EP timestamp.}  
\label{fig:unmix}
\vskip -5mm
\end{figure}

\section{EP detection for SURT}

In the multi-talker ASR scenario, EP detection is more challenging due to the presence of overlapped speech compared with the single-talker setting. More specifically, the model needs to detect the timestamp when one speaker finishes speaking while another speaker may be still speaking. To deal with this challenge within SURT, we propose to perform EP detection in the hidden feature representation space defined by the unmixing module instead of the raw feature space $X$. In this case, EP detection is performed independently for each channel as shown in Figure~\ref{fig:unmix}. As explained in Section~\ref{sec:intro}, this can be simply done by introducing a special $\langle \text{eos} \rangle$ token~\cite{chang2019joint} as a trigger for EP detection. If we denote $t_{\langle \text{eos} \rangle}$ as the ground truth endpoint timestamp and $\hat{t}_{\langle \text{eos} \rangle}$ the timestamp that the model emits the $\langle \text{eos} \rangle$ during inference, the goal of EP detection for SURT is to minimize the difference between the two for both channels. Similar to the observations in the single-talker condition~\cite{li2020towards}, without any constraint in training, the transducer-based models usually delay the emission of $\langle \text{eos} \rangle$, consequently introducing additional latency that hurts user experience in real streaming applications. To minimize the EP detection latency, we apply the latency penalty approach~\cite{li2020towards} as 
\begin{align}
\label{eq:latency}
    \log P(\langle \text{eos} \rangle \mid z_{t, u}) -= \max(0, \alpha(t - t_{\text{buffer}} - t_{\langle \text{eos} \rangle})),
\end{align}
where $t_{\text{buffer}}$ is a grace period, $t_{\langle \text{eos} \rangle}$ is the ground-truth EP timestamp and $\alpha$ is a tunable hyper-parameter. Intuitively, when computing the RNN-T loss, this approach will penalize the probability of a path that emits the $\langle \text{eos} \rangle$ beyond a grace period $t_{\text{buffer}}$ according to the reference $t_{\langle \text{eos} \rangle}$.
Another approach that works well for single-talker condition is FastEmit~\cite{yu2020fastemit}, which reduces the gradients of the blank tokens during training. However, we did not observe reduction in latency in our multi-talker ASR experiments. 

\begin{table}[t]\centering
\caption{Multi-talker E2E speech recognition results from both streaming and offline models.}
\label{tab:asr}
\footnotesize
\begin{tabular}{cccc|c}
\hline 

\hline
    System    & Size & Network    & Streaming  & WER \\ \hline
  PIT-S2S~\cite{kanda2020joint}    & 160.7M & LSTM &   \xmark   & 10.3   \\
  SOT-S2S-MBR~\cite{kanda2020minimum} & 135.6M & LSTM & \xmark & 8.3$^\dagger$ \\
    SURT~\cite{lu2021streaming} & 81M & LSTM     & \xmark  & 7.2   \\
  PIT-MS-RNN-T~\cite{sklyar2020streaming} & 80.9M & LSTM     & \cmark  & 10.2   \\
  SURT~\cite{lu2021streaming} & 81M & LSTM     & \cmark  & 10.3   \\
  SURT (this work) & 84.5M & Transformer & \cmark & 9.1 \\

\hline

\hline
\end{tabular}
\\$^\dagger$ This model used speaker inventory as an auxiliary input.
\vskip-5mm
\end{table}

\section{Experiments and Results}
\label{sec:exp}

\subsection{Dataset}
Our experiments were performed on the LibriSpeechMix dataset~\cite{kanda2020serialized}, which is a simulated overlapped audio dataset derived from the LibriSpeech corpus~\cite{panayotov2015librispeech}. We used the same protocol to simulate the training and evaluation data as in~\cite{kanda2020serialized}, and we focus on the 2-speaker overlapped speech scenario. To generate the simulated training data, for each utterance in the original LibriSpeech {\tt train\_960} set, we randomly picked another utterance from a {\it different} speaker, and mix the latter with the previous one with a random delay ranging from 0.5 seconds (large overlap) to the total length of the first utterance (no overlap). 
The number of mixed audio is the same as the number of utterances in the original LibriSpeech dataset. For both training and evaluation data, each utterance only has 2 speakers after simulation. The source code to reproduce our evaluation data is publicly available\footnote{\scriptsize \url{https://github.com/NaoyukiKanda/LibriSpeechMix}}. The ground truth EP timestamps $t_{\langle \text{eos} \rangle}$ were obtained from force alignment of the original Librispeech dataset using a hybrid ASR model trained with Kaldi~\cite{povey2011kaldi}.

\subsection{Experimental Setup}
In this work, we evaluate two network structures of SURT model that are based on RNN-T and T-T respectively. For both types of SURT model, we use the same neural network architecture for the unmixing module, i.e., 4-layer 2D CNNs for both $\text{CNN}_\text{mask}$ and $\text{CNN}_\text{enc}$ as in~\cite{lu2020streaming}. For RNN-T based SURT model, we used a 6-layer unidirectional LSTM as the audio encoder and 2-layer unidirectional LSTM as the label encoder. The number of hidden units was set to be 1024 for all LSTM layers. For T-T based SURT model, we used 18 Transformer blocks with the attention dimension as 624 and each block has 8 attention heads. In order to make T-T streamable, we applied chunk-wise attention, i.e., the self-attention operation is constrained with a chunk instead of the entire sequence~\cite{lu2020exploring, chen2021developing}. For our experiments, we set the chunk size as 25, which translates to 1 second of audio. For the latency penalty as Eq~\eqref{eq:lp}, we set $\alpha=2$ and $t_{\text{buffer}} = 3$ (corresponding to 120 milliseconds of audio) for all our experiments. 
 
In terms of features, we used the 80-dimensional log-mel fiterbanks (FBANKs) as features, which were sampled the features at 100 Hz frame rate. The CNN encoder in the unmixing module then downsampled the features to 25 Hz by max-pooling operations. For tokenization, we used 4,000 word-pieces as output tokens for RNN-T, which are generated by byte-pair encoding (BPE)~\cite{sennrich2015neural}. 
We also applied speed perturbation for data augmentation by creating two additional versions of the acoustic features with the speed ratios as 0.9 and 1.1. 

 \begin{table}[t]\centering
\caption{WER results of SURT models with EP detection and latency penalty.}
\label{tab:epd}
\footnotesize
\vskip0.15cm
\begin{tabular}{l|l|c}
\hline 

\hline
ID & Model      & WER  \\ \hline
A0 & LSTM-SURT &  10.3\\
A1 & + $\langle \text{eos} \rangle$ &   12.1\\
A2 &  $\hskip2mm$+ latency penalty &  17.5\\ \hline
B0 & Transformer-SURT & 9.1 \\
B1 & + $\langle \text{eos} \rangle$  & 10.1 \\
B2 & $\hskip2mm$+ latency penalty & 12.7 \\

\hline

\hline
\end{tabular}
\vskip-5mm
\end{table}

\begin{figure}[t]
\small
\centerline{\includegraphics[width=0.38\textwidth]{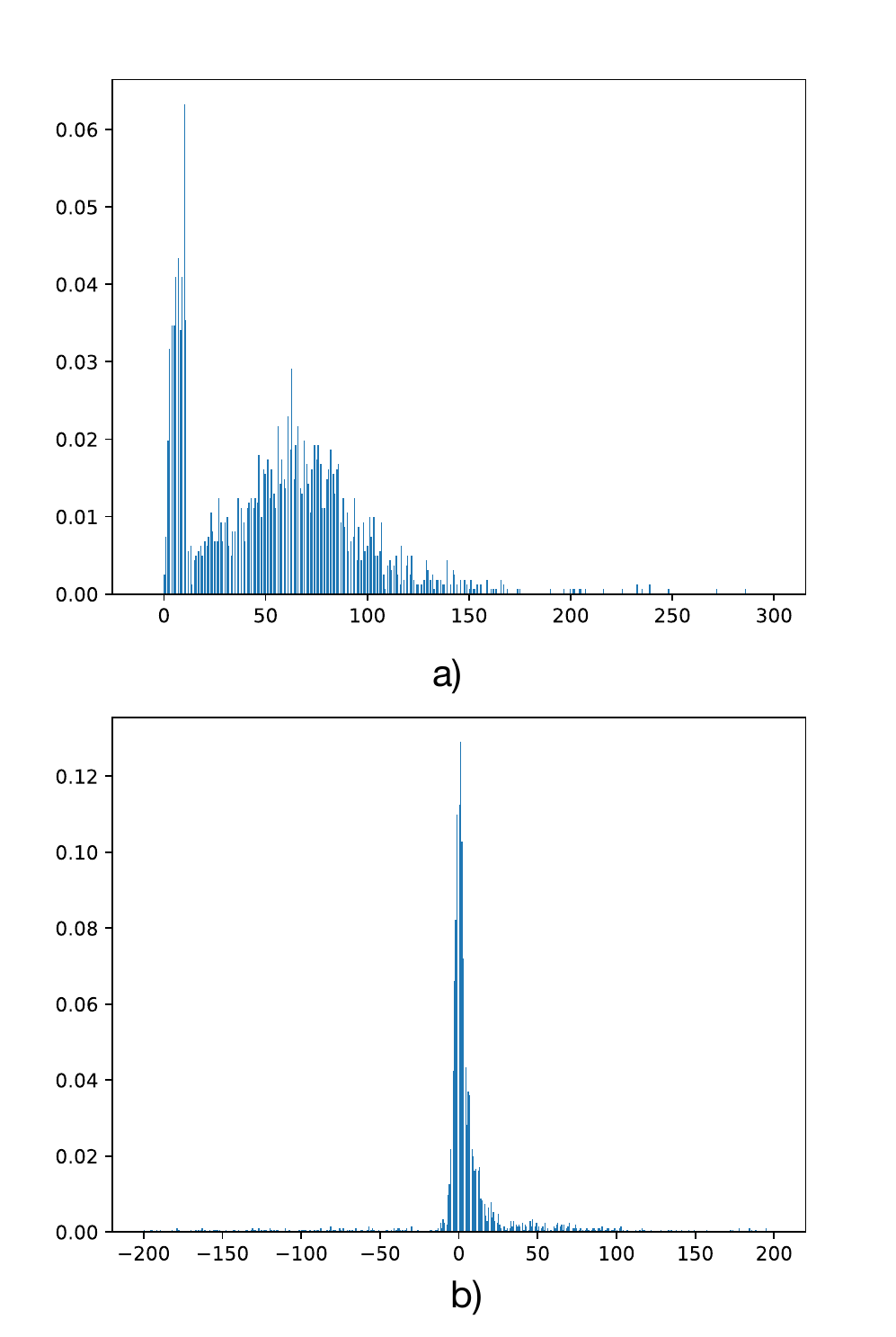}}
\vskip-4mm
\caption{The value of $\mu$ is measured by the number of frames. a) The distribution of $\mu$ from the model B1 in Table~\ref{tab:epd}. b) The distribution of $\mu$ from the model B2 in Table~\ref{tab:epd}.}  
\label{fig:eos}
\vskip -6mm
\end{figure}

\subsection{Multi-talker ASR Results}
\label{ssec:baseline}

Table~\ref{tab:asr} shows the multi-talker speech recognition word error rate (WER) results on the evaluation data derived from Librispeech {\tt test-clean} set. The table compares the SURT model results from this work with prior works using different model architectures and training objectives. In particular, PIT-S2S~\cite{kanda2020joint} and SOT-S2S-MBR~\cite{kanda2020minimum} are offline models, which are based on S2S models that are trained with PIT and serialized output training (SOT) respectively. Our previous RNN-T based SURT model~\cite{lu2021streaming} trained with HEAT achieved 10.3\% WER in the streaming scenario, and 7.2\% in the offline scenario with bidirectional LSTMs. These results compare favorably with the offline S2S models~\cite{kanda2020joint, kanda2020minimum} and the streaming RNN-T model trained with PIT~\cite{sklyar2020streaming}. In this work, we observe that T-T based SURT model improves upon the RNN-T based counterpart by over 10\% relative by only marginally increasing the model size. Note, however, the T-T based model increases the algorithmic latency by around 1 second compared with the RNN-T based model due to the chunk-wise attention mechanism. 

Table~\ref{tab:epd} show the WER results of SURT models with and without the $\langle \text{eos} \rangle$ token and the latency penalty. From our experiments, the RNN-T based SURT model is more sensitive to the $\langle \text{eos} \rangle$ token and the latency penalty compared with T-T based SURT model. As observed in previous works on single-talker RNN-T~\cite{chang2019joint, li2020towards}, integrating the $\langle \text{eos} \rangle$ into an RNN-T model usually increases the WER, because the $\langle \text{eos} \rangle$ may be emitted before the end of an utterance (a.k.a, premature EP), which will result in higher deletion error.  Compared with single-talker RNN-T model results as in~\cite{li2020towards}, the results from our SURT multi-talker models with only the $\langle \text{eos} \rangle$ token are still reasonable. However, the degradation introduced by the latency penalty is much more than expected, particularly for RNN-T based SURT model. One key reason, in fact, is due to the Librispeech dataset. We observe that some utterances are not well segmented and have considerable intermediate silence. The silence will trigger more premature EP detections, particularly for the model trained with the latency penalty as the model was encouraged to emit $\langle \text{eos} \rangle$ with silence feature input. For T-T based SURT model, the accuracy degradation from the latency penalty is much smaller, which is due to the chunk-wise self-attention operation in the Transformer network. Intuitively, if a silent region is constrained within one chunk, it will be less unlikely to trigger a premature EP detection. However, if a silent region bridges two consecutive chunks, a premature EP detection is still likely to occur. One approach to eliminate the silence issue is to perform another round of segmentation of the Librispeech data. However, we did not prefer this approach in order to make our results comparable with prior works. 

\begin{table}[t]\centering
\caption{The recall rates of EP detection from T-T based SURT model trained with and without the latency penalty, i.e., the percentage of the detected EPs with $|\mu|$ below a threshold. Note that $|\mu|$ is measured by the number of frames, and the frame rate in our experiment is 25 Hz, i.e., 5 frames correspond to 200 milliseconds. }
\label{tab:recall}
\footnotesize
\vskip0.15cm
\begin{tabular}{l|c|cccc}
\hline 

\hline
& & \multicolumn{3}{c}{$|\mu|$} \\ 
 Model     & Channel & $\le$ 5 & $\le$ 7 & $\le$ 9   \\ \hline
 w/o penalty & 1 & 0.08 & 0.13 & 0.18 \\
 w/o penalty & 2 & 0.37 & 0.72 & 0.75 \\
 w/ penalty & 1 & 0.64 & 0.72 & 0.75 \\
 w/ penalty & 2 & 0.76 & 0.85 & 0.89 \\

\hline

\hline
\end{tabular}
\vskip-3mm
\end{table}

\subsection{EP Detection Results}

Before presenting the EP detection results, we firstly denote $\mu = \hat{t}_{\langle \text{eos} \rangle} - t_{\langle \text{eos} \rangle}$. If $\mu > 0$, it corresponds to delayed EP detection, while $\mu < 0$ indicates a premature EP detection. Figure~\ref{fig:eos} shows the distribution of $\mu$ from the first channel of the T-T based SURT model with and without latency penalty. Not surprisingly, the EP detection latency from the model trained without latency penalty is very significant. In particular, there are certain cases where $\mu$ is very large, e.g., $\mu > 150$. This is mainly due to the leakage issue from the unmixing module, e.g., the features from the speaker of the 2nd channel are leaked to the first channel and vice versa. Consequently, $H_1$ and $H_2$ may have the feature representations of two speakers, which will result in a delayed EP detection and more insertion errors.  With latency penalty, we observe that the distribution of $\mu$ now centers around 0 with much smaller variance as expected. However, we still observe a small number of outliers with $|\mu| \gg 0$. As explained before, $\mu \gg 0$ indicates a leakage problem while $\mu \ll 0$ may be due to a premature EP detection.  Finally, Table~\ref{tab:recall} shows the recall rate of EP detection, i.e., the percentage of detected EPs with $|\mu|$ below a predefined threshold. The results demonstrate that the latency penalty approach can significantly improve the recall for both channels from the unmixing module. 

\section{Conclusion}
\label{sec:conc}

For most steaming ASR applications, a good user experience depends on accurate and prompt EP detections. In this work, we investigated an E2E approach for EP detection for multi-talker ASR. In particular, we focused on the SURT framework, which we recently proposed for streaming E2E multi-talker speech recognition. We studied the end-of-sentence token based approach for EP detection, and demonstrated that with the latency penalty approach during model training, we achieved significant improvement in terms of EP detection quality. We compared two SURT model architectures that are based RNN-T and T-T respectively. Our experiments showed that the T-T model with chunk-wise attention is more robust against the latency penalty during training. A limitation of current study is that our current experiments were performed on pre-defined audio segments that one segment only contains two utterances from two speakers. In the future, we shall investigate the EP detection problem with the continuous long-form audios without prior segmentation as in~\cite{raj2021continuous}.

\bibliographystyle{IEEEbib}
\bibliography{bibtex}

\end{document}